\documentclass[twocolumn,letterpaper,aps,prl,superscriptaddress,showpacs,amsmath]{revtex4-1}

\usepackage{graphicx}

\newcommand{\vc}[1]{\boldsymbol{#1}}

\begin{document}

\title{Spin-state Crossover Model for the Magnetism of Iron Pnictides}

\author{Ji\v{r}\'{\i} Chaloupka}
\affiliation{Max Planck Institute for Solid State Research,
Heisenbergstrasse 1, D-70569 Stuttgart, Germany}
\affiliation{Central European Institute of Technology,
Masaryk University, Kotl\'a\v{r}sk\'a 2, 61137 Brno, Czech Republic}

\author{Giniyat Khaliullin}
\affiliation{Max Planck Institute for Solid State Research,
Heisenbergstrasse 1, D-70569 Stuttgart, Germany}

\begin{abstract}
We propose a minimal model describing magnetic behavior of Fe-based
superconductors. The key ingredient of the model is a dynamical mixing of
quasi-degenerate spin states of Fe$^{2+}$ ion by intersite electron hoppings,
resulting in an effective local spin $S_\mathrm{eff}$. The moments $S_\mathrm{eff}$
tend to form singlet pairs, and may condense into a spin nematic phase due to
the emergent biquadratic exchange couplings. The long-range ordered part $m$
of $S_\mathrm{eff}$ varies widely, $0\leq m\leq S_\mathrm{eff}$, but magnon
spectra are universal and scale with $S_\mathrm{eff}$, resolving the puzzle
of large but fluctuating Fe-moments. Unusual temperature dependences of a
local moment and spin susceptibility are also explained. 
\end{abstract}

\date{\today}





\maketitle
Since the discovery of superconductivity (SC) in doped LaFeAsO~\cite{Kam08}, 
a number of Fe-based SC's have been found and studied~\cite{John10}. Evidence 
is mounting that quantum magnetism is an essential part of the physics of 
Fe-based SC's. However, the origin of magnetic moments and the mechanisms 
that suppress their long-range order (LRO) in favor of SC remain far 
from being well understood. 

The magnetic behavior of Fe-based SC's is unusual. The ordered moments range
from $0.1-0.4\:\mu_\mathrm{B}$, as in spin-density wave (SDW) metals like Cr,
to $1-2\:\mu_\mathrm{B}$ typical for Mott insulators, causing debates whether
the spin-Heisenberg~\cite{Yil09,Xu08,Si08,Fan08,Uhr09,Sta11} or fermionic-SDW
pictures~\cite{Maz08,Kur08,Chu08,Gra09,Kan09} are more adequate. At the same
time, irrespective to the strength or very presence of LRO, the Fe-ions
possess the fluctuating moments 
\mbox{$\sim 1-2\:\mu_\mathrm{B}$}~\cite{Gre11,Vil12}, even in apparently 
``nonmagnetic'' LiFeAs and FeSe. In fact, it was noticed early on that 
the Fe-moments, ``formed independently on fermiology''~\cite{Joha10} and 
``present all the time''~\cite{Yil09}, are instrumental to reproduce the 
measured bond-lengths and phonon spectra~\cite{Maz09,Joha10,Yil09,Rez09}. 
Recent experiments~\cite{Liu12,Zho13,Wan13} observe intense high-energy 
spin-waves that are almost independent of doping, further supporting a notion 
of local moments induced by Hund's coupling~\cite{Yin11} and
coexisting~\cite{Dai09,Kou09,Lv10} with metallic bands. 

While the formation of the local moments in multi-orbital systems is natural,
it is puzzling that these moments (residing on a simple square lattice) may
remain quantum disordered in a broad phase space despite a sizable interlayer
coupling; moreover, the Fe-pnictides are semimetals with strong tendency of
the electron-hole pairs to form SDW state, further {\it supporting} classical
LRO of the underlying moments. A fragile nature of the magnetic-LRO in
Fe-pnictides thus implies the presence of a strong quantum disorder effects,
not captured by {\it ab-initio} calculations that invariably lead to magnetic
order over an entire phase diagram. The ideas of domain wall
motion~\cite{Maz09} and local spin fluctuations~\cite{Yin11} were proposed as
a source of spin disorder, but no clear and tractable model of quantum
magnetism in Fe-based SC's has emerged to date. Here we propose such a model. 

Since Fe-pnictides are distinct among the other (Mn, Co, Ni) families, their 
unique physics should be rooted in specific features of the Fe-ion itself. 
In fact, Fe$^{2+}$ is famous for its spin-crossover~\cite{Gut04}: it may 
adopt either of $S$=$0,1,2$ states depending on orbital splitting, covalency, 
and Hund's coupling. As the ionic radius of Fe is sensitive to its spin, 
Fe-$X$ bond length ($X$ is a ligand) is also crucial. In oxides, $S$=2 is 
typical and $S$=$0,1$ occur at high pressures only~\cite{Sta08}. In
compounds with more covalent Fe-$X$ bonds ($X$=S, As, Se), $S$=0 is more
common while $S$=$1,2$ levels are higher. Here it comes the basic idea of
this Letter: when the covalency and Hund's coupling effects compete, the
many-body ground state (GS) is a {\it coherent superposition} of different
spin states intermixed by electron hoppings, resulting in an {\it average}
effective spin $S_\mathrm{eff}$ whose length depends on pressure, etc. We 
explore this dynamical spin-crossover idea, and find that: 
({\it i}) local moment $S_\mathrm{eff}$ may {\it increase} with temperature 
explaining recent data~\cite{Gre13}; 
({\it ii}) interactions between $S_\mathrm{eff}$ contain large biquadratic 
exchange, and resulting spin-nematic correlations compete with magnetic-LRO; 
({\it iii}) the ordered moment $m$ varies widely, but magnon spectra are 
universal and scale with $S_\mathrm{eff}$ as observed~\cite{Liu12,Zho13,Par12}; 
({\it iv}) singlet correlations among $S_\mathrm{eff}$ lead to the increase 
of the spin susceptibility with temperature~\cite{Kli10}.

\begin{figure}[tb] 
\includegraphics[width=8.2cm]{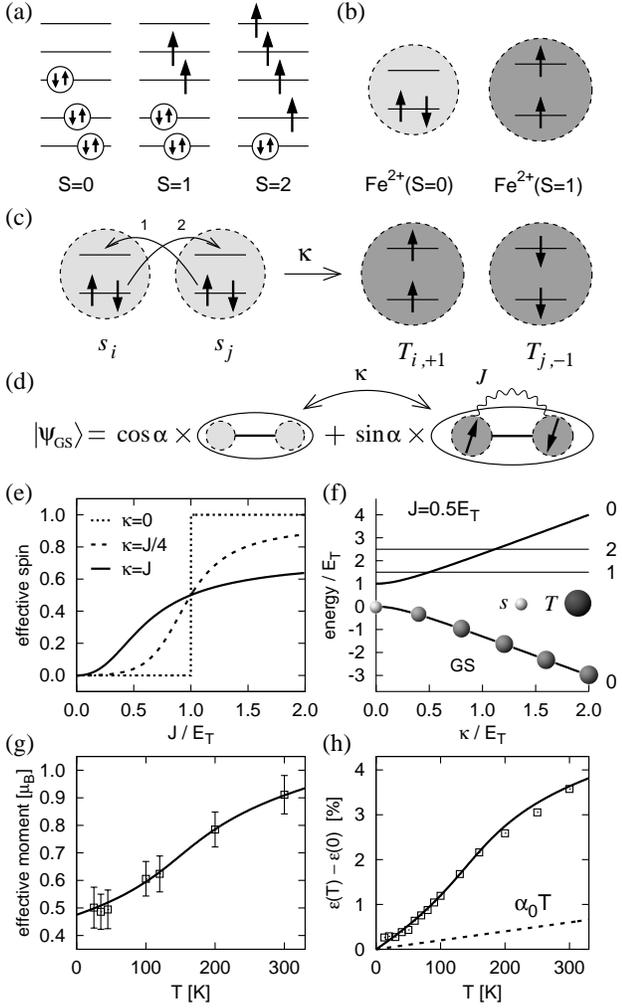} 
\caption{
(a)~Schematic view of low ($S=0$), intermediate ($S=1$), and
high ($S=2$) spin states of Fe$^{2+}(3d^6)$. 
(b)~$S=0$ and $S=1$ states differ in two electrons (out of six) occupying 
either the same or two different $t_{2g}$ orbitals. The $S=1$ state has a 
larger ionic radius.
(c)~The $\kappa$-process generating a singlet pair of $S=1$ triplets 
$T$ of two Fe$^{2+}$ ions, both originally in the $S=0$ state (denoted by $s$).
(d)~The GS wavefunction of a Fe$^{2+}$--Fe$^{2+}$ pair is a coherent 
superposition of two total-singlet states. 
(e)~Effective spin (average occupation of $S=1$ state)
depending on the ratio of the coupling $J$ between $S=1$ states and their 
energy $E_T$. 
(f)~Energy levels labeled by the total spin value of the Fe$^{2+}$--Fe$^{2+}$ 
pair. Only singlet pairs are affected by $\kappa$. With increasing $\kappa$, 
the $S=1$ states are gradually mixed into the GS.
(g)~Temperature dependence of the local magnetic moment $2n_T$, and 
(h)~the c-axis thermal expansion. Squares in (g,h) represent experimental 
data on Ca$_{0.78}$La$_{0.22}$Fe$_2$As$_2$~\cite{Gre13}. Dashed line in (h) 
is a thermal expansion excluding magnetoelastic term.
}
\label{fig:schematics}
\end{figure}

The Fe-ions in pnictides have a formal valence state Fe$^{2+}(d^6)$. Among its
possible spin states [Fig.~\ref{fig:schematics}(a)], low-spin ones are
expected to be favored; otherwise, the ordered moment would be too large and
robust. The $S=0,1$ states, ``zoomed-in'' further in
Fig.~\ref{fig:schematics}(b), are most important since they can overlap in the
many-body GS by an exchange of just two electrons between ions, see
Fig.~\ref{fig:schematics}(c). The corresponding $\kappa$-process converts
Fe($S$=0)--Fe($S$=0) pair into Fe($S$=1)--Fe($S$=1) singlet pair and vice
versa; this requires the {\it interorbital} hopping which is perfectly allowed
for $\sim 109^{\circ}$ Fe-As-Fe bonding. Basically, $\kappa$ is a part of
usual exchange process when local Hilbert space includes different spin states
$S$=0,1; hence $\kappa\sim J$. Coupling $J$ between $S$=1 triplets is
contributed also by their indirect interaction via the electron-hole Stoner
continuum and, as expected, it reduces with doping~\cite{Yar09} as the
electron-hole balance of a parent semimetal becomes no longer perfect. 

The Hamiltonian describing the above physics comprises three terms:
on-site energy $E_T$ of $S$=1 triplet $T$ relative to $S$=0 singlet $s$,
and the bond interactions $\kappa, J$: 
\begin{equation} \label{eq:model}
\mathcal{H}\!=\!E_T \!\!\sum_i \!n_{T_i} + 
\!\sum_{\langle ij\rangle} \!\!\left[-\kappa_{ij}(D^\dagger_{ij} s_i s_j \!+ 
\!\mathrm{h.c.}\!) + \!J_{ij} \vc S_i\!\cdot\!\vc S_j\right]\!.\! 
\end{equation}
The operator $D^\dagger_{ij}$ creates a singlet pair of spinfull $T$-particles 
on bond $\langle ij\rangle$. For a general spin $S$ of $T$-particles, 
$D_{ij}=\sum_M (-1)^{M+S} T_{i,+M}T_{j,-M}$ with $M=-S,\ldots,S$ denoting the 
$N=2S+1$ projections; physically, $N=3$. The constraint $n_{si}+n_{Ti}=1$ 
is implied~\cite{SM,note_orb}.

The above model rests on three specific features of Fe-pnictides/chalcogenides: 
({\it i}) spin-state flexibility of Fe$^{2+}$ that can be tuned by pressure 
increasing $E_T$, ({\it ii}) edge-sharing Fe$X_4$ tetrahedral structure
allowing ``spin-mixing'' $\kappa$-term, and ({\it iii}) semimetallic 
nature which makes $J$ values to decrease upon doping~\cite{Yar09}. 

Figure~\ref{fig:schematics}(d--f) demonstrates the behavior of 
spin-1 $T$-particles ($N=3$) on a single bond. The GS wavefunction 
$|\psi_\mathrm{GS}\rangle = \cos\alpha |A\rangle + \sin\alpha |B\rangle$
is a superposition of two singlets $A = s^\dagger_1 s^\dagger_2$ and 
$B =-\frac{1}{\sqrt 3}\sum_M (-1)^M T^\dagger_{1,M}T^\dagger_{2,-M}$, 
with the ''spin-mixing'' angle $\tan2\alpha=\sqrt{3}\kappa/(E_T-J)$. 
The GS energy $E_\mathrm{GS}=(E_T-J)-\sqrt{(E_T-J)^2+3\kappa^2 }$. 
At $\kappa=0$, there is a sudden jump [Fig.~\ref{fig:schematics}(e)] from 
$S=0$ state to $S=1$ once the $J$-energy compensates the cost of having two
$T$-particles. At finite $\kappa$, the dynamical mixing of spin states 
converts this transition into a spin-crossover, where the effective spin-length
$S_\mathrm{eff}=n_T=\sin^2\alpha$ increases gradually.
Fig.~\ref{fig:schematics}(f) shows that $\kappa$-term strongly stabilizes the
singlet pair of $T$-particles; this leads (see later) to a large biquadratic 
coupling $(\vc S_1\cdot\vc S_2)^2$ which is essential 
in Fe-pnictides~\cite{Wys11,Yar09,Yu12}. 

We are ready to show the model in action, explaining recent observation of an
unusual increase of the local moment upon warming~\cite{Gre13}. This fact is
at odds with Heisenberg and SDW pictures but easy to understand within the
spin-crossover model. Indeed, the spin-length $S_\mathrm{eff}$ may vary as a
function of $E_T$ which, in turn, is sensitive to lattice expansion; in fact,
Gretarsson {\it et al.} found that the moment value follows $c$-axis thermal
expansion $\epsilon=\delta c/c$. We add (magnetoelastic) coupling $-A\epsilon
n_T$ in Eq.~\eqref{eq:model}, affecting $E_T$ value, and evaluate $\epsilon$
and $\langle n_T\rangle_\epsilon$ self-consistently. This is done by
minimizing the elastic energy $\frac12K\epsilon^2-K\alpha_0 T\epsilon
+\frac14Q\epsilon^4$ ($\alpha_0$ is the usual thermal expansion coefficient),
together with the GS energy $E_\mathrm{GS}$ given above. This results in a
linear relation $\epsilon \simeq \alpha_0 T +\frac AK \langle
n_T\rangle_\epsilon$ between the magnetic moment ($=2n_T$) and lattice
expansion. They both strongly increase with temperature if lattice is ''soft''
enough (i.e., small $K$), as demonstrated in Fig.~\ref{fig:schematics}(g,h) by
employing the parameters $E_T-J=160\:\mathrm{meV}$, $\kappa=60\:\mathrm{meV}$,
$A=1.5\:\mathrm{eV}$, $K=4.55\:\mathrm{eV}$, $Q=250\:\mathrm{eV}$, and
$\alpha_0=0.2\times10^{-4}\:\mathrm{K}^{-1}$, providing a good fit to the
experimental data of Ref.~\cite{Gre13}. 

\begin{figure}[tb]
\includegraphics[width=8.2cm]{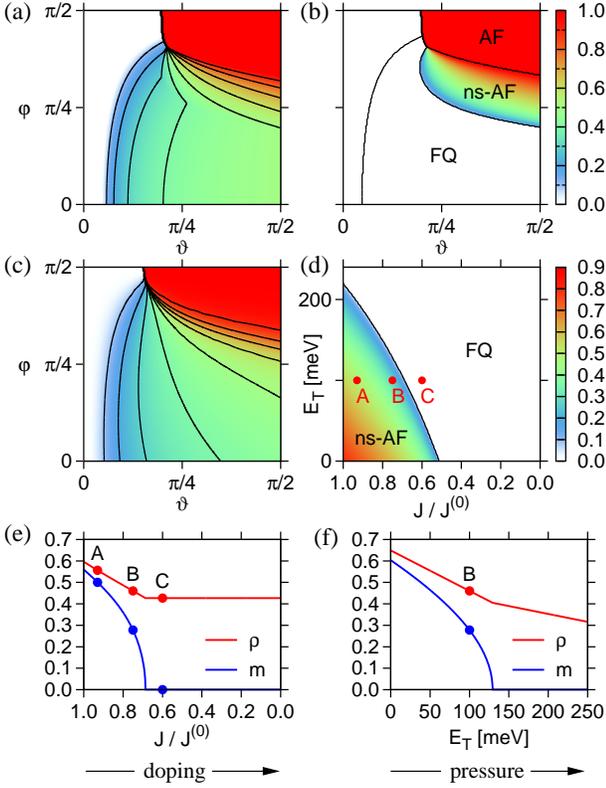}
\caption{(color online).
(a) Condensate density $\rho$ ($\equiv S_{\mathrm{eff}}$) obtained from 
Eq.~\eqref{eq:var} as a function of angles $\vartheta,\varphi$ which 
parametrize the model \eqref{eq:model} via 
$E_T=\cos\vartheta$, $\kappa_1=\sin\vartheta\cos\varphi$, and 
$J_1=\sin\vartheta\sin\varphi$. We set $\kappa_2/\kappa_1=J_2/J_1=0.7$.
(b) The ordered spin moment value $m$.
(c) $T$-occupation per site $n_T$ obtained by an exact diagonalization 
of 12-site cluster, to be compared with $\rho$ of panel (a). 
(d) The ordered moment $m$ as a function of $E_T$ and relative $J$-strength 
for fixed $\kappa_1=100\:\mathrm{meV}$, $\kappa_2=0.7\kappa_1$, 
$J_1^{(0)}=140\:\mathrm{meV}$, $J_2^{(0)}=0.7J_1^{(0)}$.
(e,f) Effective spin-length $\rho=S_{\mathrm{eff}}$ and ordered moment $m$
at the (e) $E_T=100\:\mathrm{meV}$ and (f) $J/J^{(0)}=0.75$ lines through 
the phase diagram in (d).
}
\label{fig:phasediag}
\end{figure}

Turning to collective behavior of the model, we notice first that for
$N\!\rightarrow\!\infty$ and large $\kappa$, the GS is dominated by tightly
bound singlet dimers derived from the single-bond solution. The resonance of
dimers on square-lattice plaquettes then supports a columnar state~\cite{Rea89}
breaking lattice symmetry without magnetic LRO~\cite{note_orb}. In the
opposite limit of $N=1$, the model shows a condensation of 
$T$-bosons. We found that the $N=3$ model relevant here is also unstable
towards a condensation of $T$-particles with $S=1$. This condensate hosts
interesting properties not present in a conventional Heisenberg model. We
discuss them based on the following wavefunction describing
Gutzwiller-projected condensate of spin-1 $T$-bosons:
\begin{equation}\label{eq:var}
|\Psi\rangle = \prod_i \Bigl[\sqrt{1-\rho}\;s^\dagger_i
+\sqrt{\rho}\,\sum_{\alpha=x,y,z} d^\ast_{\alpha i} T^\dagger_{\alpha i} 
\Bigr]\,|\mathrm{vac}\rangle \;,
\end{equation}
where $\rho\in[0,1]$ is the condensate density to be understood as the 
effective spin-length $S_\mathrm{eff}$. The complex unit vectors 
$\vc d_i=\vc u_i+i\vc v_i$ ($u_i^2+v_i^2=1$) determine the spin structure of 
the condensate in terms of the coherent states of spin-1 \cite{Iva03,Lau06} 
corresponding to $T_x=(T_{+1}-T_{-1})/\sqrt{2}i$, 
$T_y=(T_{+1}+T_{-1})/\sqrt{2}$, $T_z=iT_0$. The GS phase diagram obtained by 
minimizing $\langle\Psi|\mathcal{H}|\Psi\rangle$ and cross-checked 
by an exact diagonalization on a small cluster is presented in 
Fig.~\ref{fig:phasediag}. We have included nearest-neighbor (NN) and 
next-NN interactions and fixed their ratio at $J_2/J_1=\kappa_2/\kappa_1=0.7$, 
reflecting large next-NN overlap via As ions. Like in $J_1-J_2$ model, 
this ratio decides between $(\pi,\pi)$ and $(\pi,0)$ order. 
Fig.~\ref{fig:phasediag}(a,b) contains, apart from a disordered 
(uncondensed) phase ($\rho=0$) at small $\kappa, J$, three 
distinct phases depending on $\kappa/E_T$ and $J/E_T$ values: 
(\textit{i})~Ferroquadrupolar (FQ) phase with $\vc u_i=\vc u$ 
and $\vc v_i=0$. This phase has zero magnetization and 
is characterized by the quadrupolar order parameter 
$\langle S^\alpha S^\beta - \frac13 S^2\delta_{\alpha\beta} \rangle =
\rho\,( \frac13\delta_{\alpha\beta} - u_\alpha u_\beta)$ with $\vc u$ 
playing the role of the \textit{director}~\cite{Lau06}. 
This state, often referred to as {\it spin-nematic}, appears 
in biquadratic-exchange~\cite{Iva03,Lau06,Tsu06,Har02} and 
optical lattice models~\cite{Dem02,Yip03,Pue08,Ser11}.
(\textit{ii})~Non-saturated antiferromagnetic (ns-AF) phase with stripy
magnetic order, specified by $\vc u_i=(0,0,u)$ and 
$\vc v_i=(0,v,0)\,\mathrm{e}^{i\vc Q\cdot\vc R_i}$ with $\vc Q=(\pi,0)$. 
The LRO-moment $\langle\vc S\rangle$ given by $m=2\rho uv$ can 
take values from 0 to $S_\mathrm{eff}=\rho$.
(\textit{iii})~Saturated antiferromagnet (AF) with the same $\vc Q$ vector,
but now with $u=v=1/\sqrt{2}$ and $m=S_\mathrm{eff}=1$.

The part of the phase diagram relevant to pnictides is shown in 
Fig.~\ref{fig:phasediag}(d). The decrease of $J$ is associated with doping 
that changes the nesting conditions~\cite{Yar09}, while the increase of 
$E_T$ is related to external/chemical pressure. Fig.~\ref{fig:phasediag}(e,f) 
shows that the LRO-moment $m$ quickly vanishes as $J$ ($E_T$) values 
decrease (increase); however, the spin-length $S_\mathrm{eff}=\rho$ remains 
almost constant ($\sim 1/2$), corresponding to a fluctuating magnetic moment 
$\sim 1\:\mu_\mathrm{B}$. This quantum state is driven by $\kappa$-process 
which generates the spin-1 states in a form of singlet pairs. 

\begin{figure}[tb]
\includegraphics[width=8.2cm]{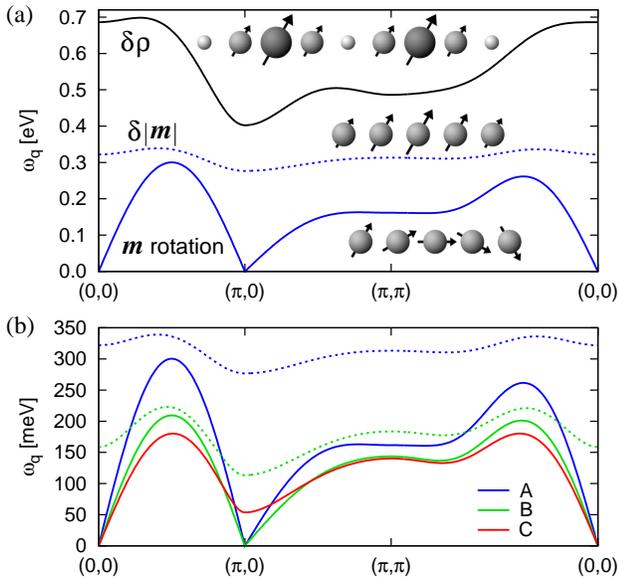}
\caption{(color online).
(a) Dispersion of the condensate density ($\delta\rho$, solid-black) and the 
ordered moment-length ($\delta |{\bf m}|$, dotted-blue) fluctuations, and the 
magnon dispersion (solid-blue), at the point $A$ in the phase diagram of 
Fig.~\ref{fig:phasediag}(d). All three modes are active in resonant 
x-ray scattering, and the latter two in neutron scattering. 
(b) Evolution of the magnetic excitations going from FQ to the ns-AF phase
[$C\to B\to A$ in Fig.~\ref{fig:phasediag}(d)]. Two-fold degenerate 
quadrupole-waves ($C$) split into the magnon (solid lines) and the 
$\delta|{\bf m}|$ mode (dotted lines). The latter represents oscillations 
between the nematic and magnetic orderings and is gapful. 
}
\label{fig:excit}
\end{figure}

We consider now the excitation spectrum. It is convenient to separate 
fast (density) and slow (spin) fluctuations. We introduce pseudospin 
$\tau=1/2$ indicating the presence of a $T$-particle, and a vector field 
$\vc d$ defining the spin-1 operator as 
$\vc S = -i(\vc d^\dagger\times\vc d)$. The resulting Hamiltonian
\begin{multline}\label{eq:Htaud}
\mathcal{H}= E_T \sum_i \left(\tfrac12-\tau_i^z\right)
-\sum_{\langle ij\rangle} \kappa_{ij} \,
( \tau^+_i\tau^+_j \,\vc d_i\cdot\vc d_j + \mathrm{h.c.}) \\
-\sum_{\langle ij\rangle} J_{ij} 
\left(\tfrac12-\tau_i^z\right) \left(\tfrac12-\tau_j^z\right)
(\vc d^\dagger_i\times\vc d_i)\cdot(\vc d^\dagger_j\times\vc d_j) \; 
\end{multline}
is decoupled on a mean-field level. The condensate spin dynamics is
then given by $O(3)$-symmetric Hamiltonian 
\begin{equation}\label{eq:Hd}
\mathcal{H}_d\!=\!
-\!\sum_{\langle ij\rangle} \tilde\kappa_{ij} (\vc d_i\cdot\vc d_j + \mathrm{h.c.}\!) 
-\sum_{\langle ij\rangle} \tilde J_{ij} 
(\vc d^\dagger_i\times\vc d_i)\cdot(\vc d^\dagger_j\times\vc d_j)
\end{equation}
with the renormalized
$\tilde{\kappa}_{ij}= \kappa_{ij}\langle \tau^+_i\tau^+_j\rangle 
\approx \kappa_{ij} (1-\rho)\rho$ and $\tilde{J}_{ij}\approx J_{ij} \rho^2$. 
The excitations are found by introducing $a$, $b$, $c$ bosons according to 
$\vc d =(d_x,d_y,d_z)= ( a, u\,b-iv\,\mathrm{e}^{i\vc Q\cdot\vc R}\, c ,
-iv\,\mathrm{e}^{i\vc Q\cdot\vc R}\, b +u\,c )$, and replacing the condensed 
one as $c,c^\dagger\!\rightarrow\!\sqrt{1-n_a-n_b}$. The resulting $(a,b)$ 
Hamiltonian is solved by the Bogoliubov transformation. A similar approach 
is used for the $\tau$-sector describing the condensate density fluctuations 
$\delta\rho =\delta S_{\mathrm{eff}}$. 

Shown in Fig.~\ref{fig:excit} is the excitation spectra for several points of
the phase diagram. The spin-length fluctuations $\delta S_{\mathrm{eff}}$ are
high in energy. Fig.~\ref{fig:excit}(b) focuses on the magnetic excitations.
In the FQ phase, quadrupole/magnetic modes are degenerate and gapless at 
$\vc q=0$. As the AF phase is approached, the gap at $\vc Q$ decreases, and 
closes upon entering the magnetic phase. However, the higher energy magnons 
(which scale with $S_{\mathrm{eff}}$) are not much affected by transition, 
apart from getting (softer) harder in a (dis)ordered phase; this explains 
the persistence of well-defined high-energy magnons into nonmagnetic
phases~\cite{Liu12,Zho13}.

The magnetic modes in Fig.~\ref{fig:excit}(b) resemble excitations of
bilinear-biquadratic spin model~\cite{Lau06}. In fact, the dispersion in FQ
phase can be \textit{exactly} reproduced~\cite{note_bqH} from an effective
spin-1 model $\sum_{\langle ij\rangle} \tilde J_{ij} \vc S_i\cdot \vc S_j
-\tilde\kappa_{ij} (\vc S_i\cdot \vc S_j)^2$, with $\tilde J$ and $\tilde
\kappa$ given above. A large biquadratic coupling was indeed found to account
for many observations in Fe-pnictides \cite{Wys11,Yar09,Sta11}. We note
however, that this model possesses FQ and AF phases only and misses the ns-AF
phase, where the ordered moment is reduced already at the classical level;
also, it does not contain the key notion of the original model,
i.e., formation of the effective spin $S_\mathrm{eff}$ and its fluctuations. 

\begin{figure}[tb]
\includegraphics[width=8.2cm]{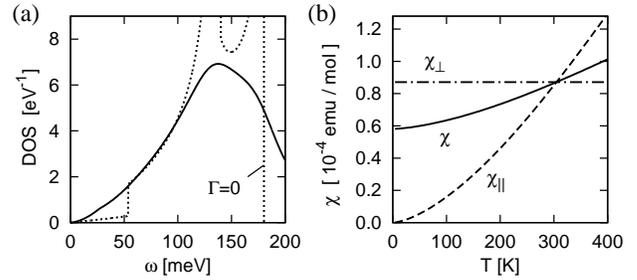}
\caption{
(a) Density of states of the magnetic excitations calculated for the point 
$C$ of Fig.~\ref{fig:phasediag}(d). We included the damping (e.g., due to
coupling to the Stoner continuum) in a form
$\Gamma(\omega)=\mathrm{min}(\omega,\Gamma)$ with $\Gamma=\omega_{\vc Q}/2$.
The result with $\Gamma=0$ is shown for comparison.
(b) Temperature dependence of the uniform susceptibility $\chi$. The 
components $\chi_\parallel$ ($\chi_\perp$) parallel (perpendicular) to the
local director $\vc u$ are also shown.
}
\label{fig:chi0}
\end{figure}

Singlet correlations inherent to the model may also lead to increase
of the paramagnetic susceptibility $\chi(T)$ with temperature~\cite{Kli10}.
Considering nonmagnetic phase, we find that for the field parallel to the
director $\vc u$, $\chi$ is temperature dependent, $\chi_\parallel =
\frac{1}{2T}\int\mathrm{d}\omega\mathcal{N}(\omega)\sinh^{-2}\frac{\omega}{2T}$,
where $\mathcal{N}(\omega)=\sum_{\vc q}\delta(\omega-\omega_{\vc q})$ is the
density of states (DOS) of magnetic excitations, while $\chi_\perp$ is 
constant. The average $\chi=(\chi_\parallel+2\chi_\perp)/3$ (with additional 
factor of $4\rho^2\mu_\mathrm{B}^2N_A$) gives the measured $\chi(T)$, assuming 
slow rotations of the director. The DOS shown in Fig.~\ref{fig:chi0}(a) is
contributed mainly by the regions around $(\pi,0)$ and $(0,\pi)$ hosting AF
correlations. The corresponding thermal excitations lead to the increase of
$\chi$ [Fig.~\ref{fig:chi0}(b)].

To conclude, we proposed the model describing quantum magnetism of
Fe-pnictides. Their universal magnetic spectra, wide-range variations of the
LRO-moments, emergent biquadratic-spin couplings are explained. The model
stands also on its own: extending the Heisenberg models to the case of
``mixed-spin'' ions, it represents a novel many-body problem. Of a particular
interest is the effect of band fermions which should have a strong impact on
low energy dynamics of the model, e.g., converting the $\vc q=0$ Goldstone
modes into overdamped spin-nematic fluctuations. Understanding the effects of
coupling between local moments and band fermions, including implications for
SC, should be the next step towards a complete theory of Fe-pnictides. 

J.C. acknowledges support by the Alexander von Humboldt Foundation, 
ERDF under project CEITEC (CZ.1.05/1.1.00/02.0068) 
and EC 7$^\text{th}$ Framework Programme (286154/SYLICA).


\newpage
\onecolumngrid

\renewcommand{\theequation}{S\arabic{equation}}
\renewcommand{\thefigure}{S\arabic{figure}} 

\setcounter{equation}{0}
\setcounter{figure}{0}

\begin{center}
Supplemental Material for 
\vskip 2mm
{\bf\large Spin-state crossover model for the magnetism of iron pnictides}
\vskip 4mm
Ji\v{r}\'{\i} Chaloupka$^{1,2}$ and Giniyat Khaliullin$^1$
\vskip 2mm
\em\footnotesize
$^1$~Max Planck Institute for Solid State Research,
Heisenbergstrasse 1, D-70569 Stuttgart, Germany \\
$^2$ Central European Institute of Technology,
Masaryk University, Kotl\'a\v{r}sk\'a 2, 61137 Brno, Czech Republic
\vskip 3mm
\end{center}

\vskip 3mm
Here we analyze two-orbital Hubbard model in the regime of large Hund's 
coupling and large interorbital hopping, and explicitly demonstrate the 
emergence of the effective model proposed in the main paper. We also provide 
estimates of the model parameters in terms of the microscopic parameters 
of the Hubbard model.

Based on the ''orbital-differentiation'' mechanism -- which is particularly 
pronounced in multiorbital systems with large Hund's coupling 
(see Ref.~\cite{Geo12} for recent discussion) -- we assume a coexistence 
of strongly correlated orbitals (hosting magnetic moments) and more itinerant 
bands (responsible for the charge transport and Fermi-surface related physics). 
For the Fe-pnictide/chalcogenide families, a minimal model for the
''magnetic'' sector is a two-orbital Hubbard model which may accommodate  
magnetic moments ranging from zero to $2\:\mu_\mathrm{B}$ per Fe-ion, depending 
on the parameter regime. This possible moment-window is what observed in 
Fe-pnictide/chalcogenides~\cite{Gre11} (and also consistent with the model of 
the main text). We assume that these two orbitals (labeled $a$ and $b$ below) 
are populated by two electrons per site on average, while the remaining four
electrons out of Fe-$d^6$ configuration form a semimetallic band structure.  
The itinerant bands are not a prime source of magnetic moments but, as 
noticed in the main text, we keep in mind that they may mediate the  
interactions between local moments and hence support their long-range 
order~\cite{Joh10,Joh09}. 

Let us focus now on the ''magnetic'' sector, i.e. two-orbital Hubbard 
Hamiltonian. As usual, it comprises two parts, local interactions and 
intersite hoppings:
$\mathcal{H}=\mathcal{H}_\mathrm{loc} + \mathcal{H}_\mathrm{kin}$. 
Its local part includes the crystal field splitting $\Delta$ between $a$ 
and $b$ orbitals (their precise structure in terms of original $d$ states 
is not essential here) and local correlations:
\begin{equation}\label{Hloc}
\mathcal{H}_\mathrm{loc} = \frac{\Delta}2 \sum_{i} (n_{ib} - n_{ia})
+ U \sum_{i,\gamma=a,b} n_{i\gamma\uparrow} n_{i\gamma\downarrow}
+ \sum_{i}\left[ U'- J_H \Bigl(2\vc S_{ia} \cdot \vc S_{ib}
+ \tfrac12\Bigr)\right] n_{ia} n_{ib} \;.
\end{equation}
The local pair-hopping term is neglected, and the relation $U'=U-2J_H$ between
inter- and intra-orbital Coulomb interactions will be used. 
The kinetic term $\mathcal{H}_\mathrm{kin}$ of the Hamiltonian contains
the intersite hopping of both intra- and inter-orbital character
\begin{equation}\label{Hhop}
\mathcal{H}_\mathrm{kin} = -t \sum_{\langle ij\rangle, \sigma} \left(
a^\dagger_{i\sigma} a^{\phantom{\dagger}}_{j\sigma} + 
b^\dagger_{i\sigma} b^{\phantom{\dagger}}_{j\sigma} + \mathrm{h.c.} \right) 
-\tilde{t} \sum_{\langle ij\rangle, \sigma} \left(
a^\dagger_{i\sigma} b^{\phantom{\dagger}}_{j\sigma} + 
b^\dagger_{i\sigma} a^{\phantom{\dagger}}_{j\sigma} + \mathrm{h.c.} \right)
\;.
\end{equation}
Similar model was recently considered in Ref.~\cite{Kun11} to address the 
spin-transition physics in cobaltates. The key difference of our model is the
presence of interorbital hopping $\tilde{t}$, which converts the transitions 
found in Ref.~\cite{Kun11} into a smooth spin-crossover such that the ground 
state magnetic moment length (not long-range order parameter!) may acquire 
any value from zero to $2\:\mu_\mathrm{B}$. 

Our aim is to obtain the model Hamiltonian of the main paper as 
an effective low-energy Hamiltonian resulting from  
$\mathcal{H} = \mathcal{H}_\mathrm{loc} + \mathcal{H}_\mathrm{kin}$
in the appropriate regime of parameters $\Delta, J_H,$ etc. 
This is achieved by a standard procedure -- we select the relevant 
$d_i^2-d_j^2$ bond states from the eigenbasis of $\mathcal{H}_\mathrm{loc}$
and obtain effective interactions on the bonds by eliminating 
$\mathcal{H}_\mathrm{kin}$ perturbatively, employing the low-energy 
$d_i^3-d_j^1$ and $d_i^1-d_j^3$ configurations as the intermediate states. 
To check the validity of this approach, the exact eigenstates of 
$\mathcal{H}$ on a single bond are calculated and the results 
compared with those of the effective Hamiltonian $\mathcal{H}_\mathrm{eff}$ 
we have derived.

In the spin-crossover regime discussed in the main paper, large Hund's coupling 
nearly compensates the crystal field splitting (i.e., $\Delta \sim 3J_H$) and 
makes the on-site singlet 
$|s\rangle = a^\dagger_\uparrow a^\dagger_\downarrow\, |\;\rangle$
and the three triplet states
$|T_{+1}\rangle = a^\dagger_\uparrow b^\dagger_\uparrow\, |\;\rangle$,
$|T_{0}\rangle =\frac1{\sqrt{2}} 
(a^\dagger_\uparrow b^\dagger_\downarrow
+a^\dagger_\downarrow b^\dagger_\uparrow)\, |\;\rangle$, and
$|T_{-1}\rangle = a^\dagger_\downarrow b^\dagger_\downarrow\, |\;\rangle$
quasidegenerate. These states thus form the relevant low-energy sector while 
the other states (such as 
$a^\dagger_\uparrow b^\dagger_\downarrow\,|\;\rangle$) are much higher 
in energy and can be ignored. 

\begin{figure}[t]
\begin{center}
\includegraphics[width=15.5cm]{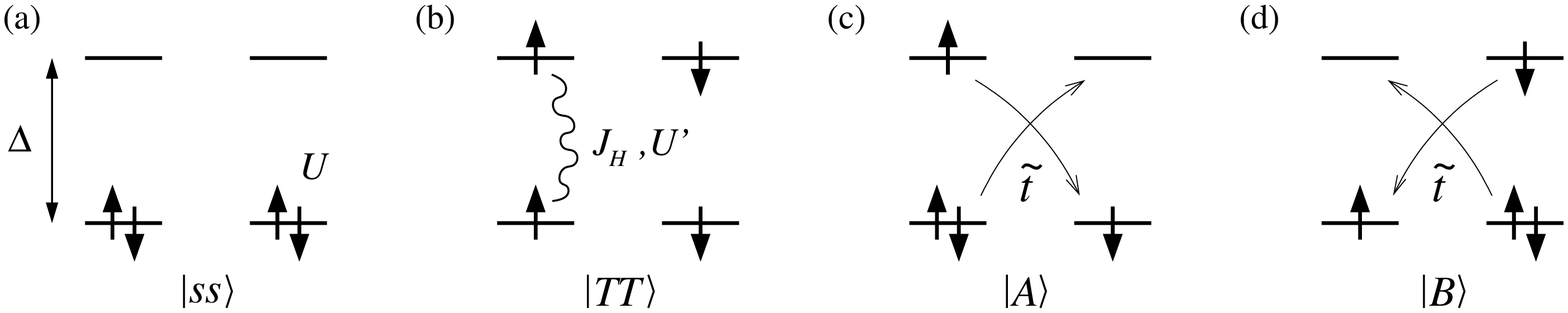}
\end{center}
\vskip -5mm
\caption{\small Basis states dominating the groundstate of the Hubbard
model in the discussed spin-crossover regime. All the states have zero total 
spin. The two $d_i^2-d_j^2$ configurations $|ss\rangle$ and $|TT\rangle$ are 
shown in panels (a) and (b), respectively, together with a schematic depiction 
of the microscopic parameters. The $d_i^3-d_j^1$ configuration $|A\rangle$ and 
$d_i^1-d_j^3$ configuration $|B\rangle$, shown in panels (c) and (d), 
respectively, are connected to the $d_i^2-d_j^2$ configurations by virtue of 
the interorbital hoppings $\tilde{t}_{ab}$ or $\tilde{t}_{ba}$ indicated 
by arrows.}
\label{fig:states}
\end{figure}

To be able to extract the effective Hamiltonian on a bond, it is convenient 
to consider the subspaces with total spin $S_\mathrm{tot}=0,1,2$ separately. 
In $S_\mathrm{tot}=0$ sector, the relevant bond states are the two 
$d_i^2-d_j^2$ configurations depicted in Fig.~\ref{fig:states}(a, b): 
$|ss\rangle=|s\rangle_i|s\rangle_j$ with the local energy $E_{ss}=2U-2\Delta$, 
and $|TT\rangle=\frac1{\sqrt{3}}\Bigl(|T_{+1}\rangle_i|T_{-1}\rangle_j -
|T_{0}\rangle_i|T_{0}\rangle_j+|T_{-1}\rangle_i|T_{+1}\rangle_j\Bigr)$ with
the local energy $E_{TT}=E_{ss}+2(\Delta-3J_H)$. The bond interaction
originates from virtual processes employing as the intermediate states mainly 
the low-lying $d_i^3-d_j^1$ and $d_i^1-d_j^3$ configurations presented in
Fig.~\ref{fig:states}(c, d). They are denoted as $|A\rangle$ and 
$|B\rangle$ and their local energy amounts to $E_A=E_B=E_{ss}+U'+\Delta-3J_H$.
The other bond states have a negligible contribution to the groundstate, due
to their high energy or due to kinematic (no hopping) reasons.
The lowest state in the $S_\mathrm{tot}=1$ sector is composed of a pair of
on-site triplets $|T\rangle$ and states analogous to $|A\rangle$ and
$|B\rangle$ but having total spin one.
Finally, the only states in the $S_\mathrm{tot}=2$ are the combinations
of two on-site triplets. These states are unaffected by hopping.

\begin{figure}[b]
\begin{center}
\includegraphics[width=14cm]{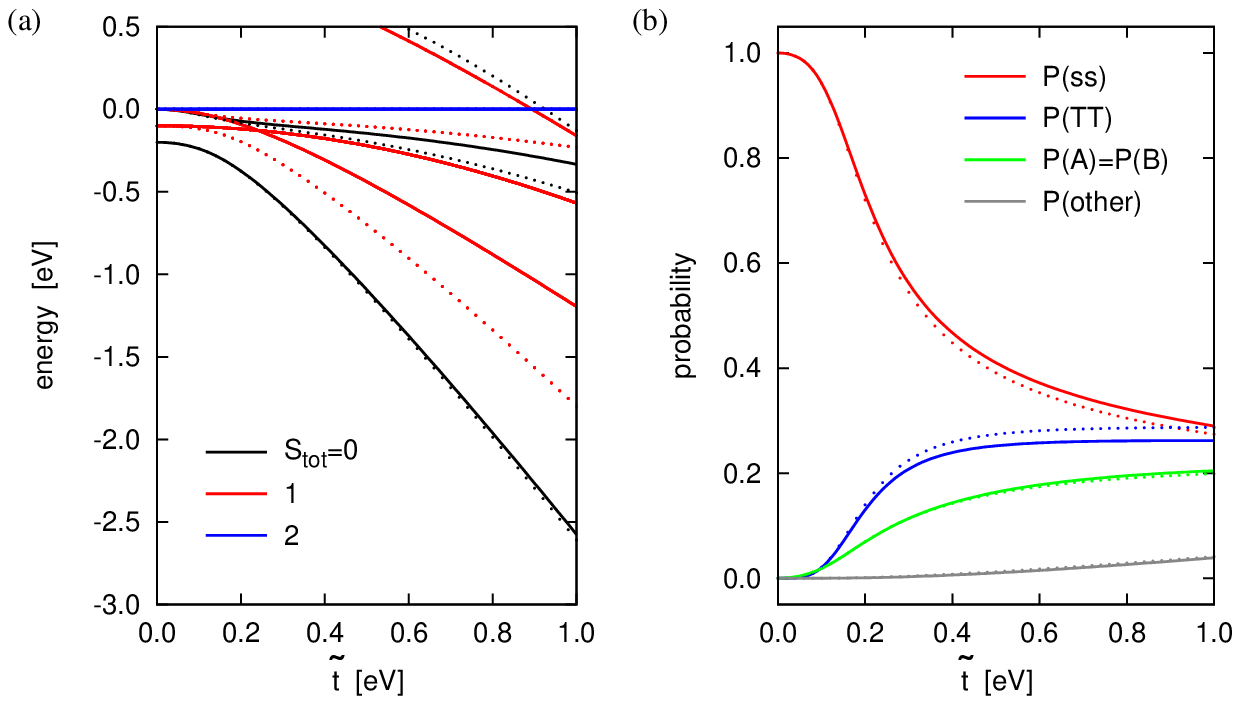}
\end{center}
\vskip -5mm
\caption{\small (a) Exact energy levels on the bond as a function of 
interorbital hopping $\tilde{t}$ for $U=3\:\mathrm{eV}$, 
$J_H=1\:\mathrm{eV}$, $\Delta-3J_H =0.1\:\mathrm{eV}$, and
$t=0$ (solid lines) and $t=0.5\,\tilde{t}$ (dotted lines). 
States with different values of the total spin are distinguished by color.
(b) Probabilities of selected basis states $|ss\rangle$, $|TT\rangle$,
$|A\rangle$, and $|B\rangle$ in the groundstate. The other states ignored 
in the effective model derivation have a negligible total weight [see
corresponding $P$(other) curve]. For a moderate value of $t$, the energy 
and the composition of the groundstate remains practically unaffected. 
}
\label{fig:exact}
\end{figure}

The validity of the above classification of low-energy levels of Hubbard model 
is demonstrated in Fig.~\ref{fig:exact} showing the results of an exact 
diagonalization of full two-orbital model $\mathcal{H}$ on a single bond. 
We consider a representative set of parameters $\Delta$ and $J_H$ such that 
a spin-crossover regime, where the on-site singlet and triplet states are 
quasidegenerate, is realized: $\Delta-3J_H =0.1\:\mathrm{eV}$. 
Focusing on $S_\mathrm{tot}=0$ sector in Fig.~\ref{fig:exact}(b), we can observe
that with increasing $\tilde{t}$, the state $|TT\rangle$ gets gradually 
involved into the groundstate, which becomes a mixture of 
$|ss\rangle$, $|TT\rangle$ and the higher energy states 
$|A\rangle$, $|B\rangle$ serving as the intermediate states for the 
$\kappa$-processes. The contribution of the other states, which are neglected 
in our derivation of the effective Hamiltonian below, is indeed negligible. 

Having selected our basis states and evaluated their local energy, we proceed 
now by incorporating the intersite hopping within this basis. First, the 
initial Hamiltonian $\mathcal{H}$ is projected to the selected subspace of 
total spin $S_\mathrm{tot}$ and denoted accordingly as $\mathcal{H}^{(S)}$ 
(where $S=0,1,2$). In the next step, the intermediate states are eliminated 
from $\mathcal{H}^{(S)}$-matrix by perturbation theory. After these 
steps, we will obtain an effective Hamiltonian $\mathcal{H}^{(S)}_\mathrm{eff}$ 
that operates within $d_i^2-d_j^2$ configuration alone, and compare it with 
the model Hamiltonian $\mathcal{H}_\mathrm{model}$ used in the main paper. 

In the most interesting $S_\mathrm{tot}=0$ subspace, after the elimination of
intermediate states, the Hamiltonian $\mathcal{H}$ projected to the subspace 
spanned by $|ss\rangle$, $|TT\rangle$, $|A\rangle$, $|B\rangle$ states 
\begin{equation}\label{Hsub44}
\mathcal{H}^{(0)} = \begin{pmatrix}
E_{ss} & 0 & -\sqrt{2}\,\tilde{t} & -\sqrt{2}\,\tilde{t} \\
0 & E_{TT} & -\sqrt{\tfrac32}\,\tilde{t} & -\sqrt{\tfrac32}\,\tilde{t} \\
-\sqrt{2}\,\tilde{t} & -\sqrt{\tfrac32}\,\tilde{t} & E_{A} & 0 \\
-\sqrt{2}\,\tilde{t} & -\sqrt{\tfrac32}\,\tilde{t} & 0 & E_{B} 
\end{pmatrix}
\quad
\text{becomes}
\quad
\mathcal{H}^{(0)}_\mathrm{eff}
= \begin{pmatrix}
E_{ss} -\dfrac{4\,\tilde{t}^{\,2}}\varepsilon &
-\dfrac{2\sqrt{3}\,\tilde{t}^{\,2}}\varepsilon \\
\rule{0pt}{9.5mm}
-\dfrac{2\sqrt{3}\,\tilde{t}^{\,2}}\varepsilon & 
E_{TT} -\dfrac{3\,\tilde{t}^{\,2}}\varepsilon 
\end{pmatrix}
\end{equation}
operating now within the $|ss\rangle$ and $|TT\rangle$ singlet states of 
$d_i^2-d_j^2$ configuration. Here, $\varepsilon=E_A-E=E_B-E$ denotes the 
excitation energy. In the second order perturbation theory $E=E_{ss}$, but 
by diagonalizing the energy dependent $\mathcal{H}^{(0)}_\mathrm{eff}$ 
~self-consistently, one can exactly reproduce the groundstate energy
and the ratio of $|ss\rangle$ and $|TT\rangle$ coefficients 
obtained by diagonalizing the original matrix $\mathcal{H}^{(0)}$.
In the following, we therefore take $E=E_\mathrm{GS}$ with $E_\mathrm{GS}$
being the groundstate energy of $\mathcal{H}_\mathrm{eff}^{(0)}$.

Using the same procedure, the pairs of local triplets $T$ of total spin
$S_\mathrm{tot}=1$ obtain an energy 
$\mathcal{H}^{(1)}_\mathrm{eff}=E_1=E_{TT}-2\tilde{t}^{\,2}/\varepsilon'$
with $\varepsilon'={E_A-E_1}$ being the excitation energy, and those of total 
spin $S_\mathrm{tot}=2$ remain at an energy $\mathcal{H}^{(2)}_\mathrm{eff}=E_{TT}$.

\begin{figure}[htb]
\begin{center}
\vskip 8mm
\includegraphics[width=14cm]{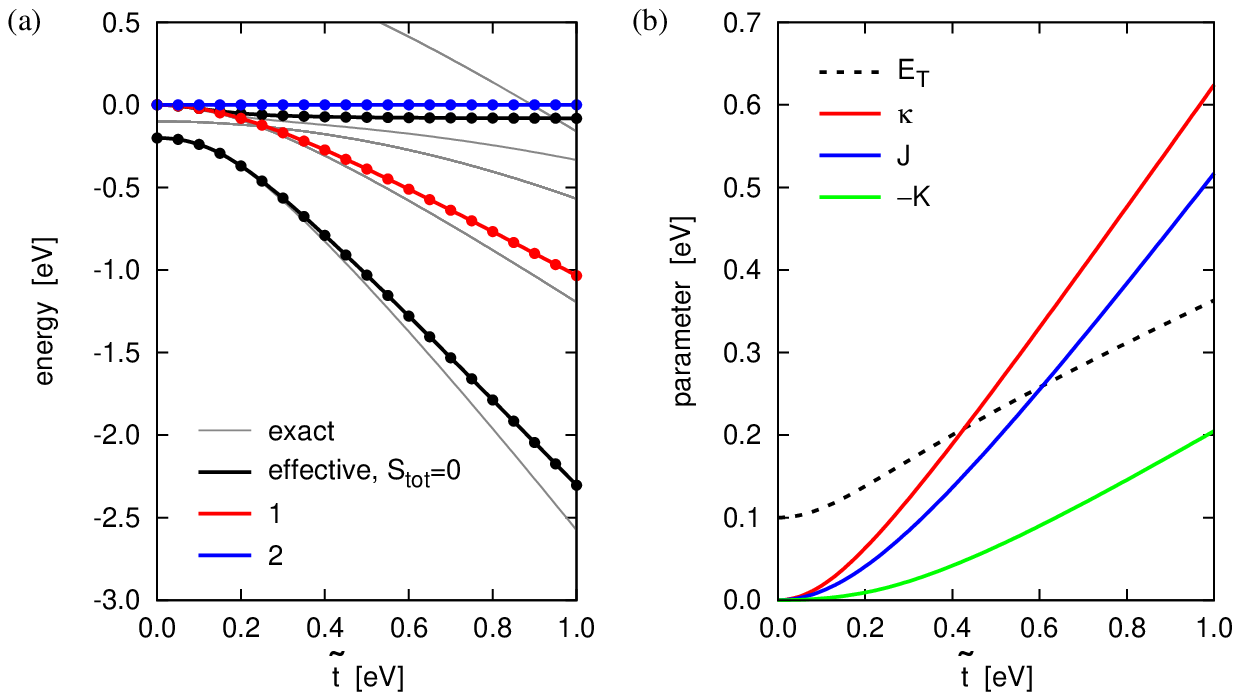}
\end{center}
\vskip -5mm
\caption{\small (a) Energy levels resulting from the diagonalization of 
$\mathcal{H}_\mathrm{eff}$ compared to the exact levels of the original 
Hubbard Hamiltonian.
The same parameters as in Fig.~\ref{fig:exact} are used and $t=0$.
(b) Values of the effective model parameters as functions of interorbital
hopping $\tilde{t}$.
}
\label{fig:model}
\end{figure}

The effective Hamiltonian $\mathcal{H}_\mathrm{eff}$ can now be exactly mapped 
to the model Hamiltonian $\mathcal{H}_\mathrm{model}$ proposed in the main 
paper. For a single bond, using the same notations, the corresponding matrix 
elements of $\mathcal{H}_\mathrm{model}$ read as
\begin{equation}\label{Hmodel}
\mathcal{H}^{(0)}_\mathrm{model}=
\begin{pmatrix}
0 & -\sqrt{3}\,\kappa \\
-\sqrt{3}\,\kappa & 2E_T-2J-4K
\end{pmatrix} \;,
\quad\mathcal{H}^{(1)}_\mathrm{model}=
2E_T-J-K \;,
\quad\mathcal{H}^{(2)}_\mathrm{model}=
2E_T+J-K \;.
\end{equation}
To make the correspondence between $\mathcal{H}_\mathrm{eff}$ and 
$\mathcal{H}_\mathrm{model}$ matrices complete, we had to include small 
biquadratic exchange $-K(\vc S_1\cdot\vc S_2)^2$. The term-by-term comparison 
of the matrix elements of $\mathcal{H}_\mathrm{eff}^{(S)}$ and 
$\mathcal{H}_\mathrm{model}^{(S)}$ yields the following values of the model 
parameters
\begin{equation}\label{kappaJ}
\kappa = \frac{2\tilde{t}^{\,2}}{\varepsilon} \;,\qquad
J = \frac{\tilde{t}^{\,2}}{\varepsilon'} \;,\qquad
K = \tilde{t}^{\,2} \left(\frac{1}{\varepsilon}-\frac{1}{\varepsilon'}\right) 
\;,\qquad
E_T = (\Delta-3J_H) + \tilde{t}^{\,2} \left(\frac{5}{2\varepsilon}-
\frac{1}{\varepsilon'}\right) \;.
\end{equation}
As evidenced by Fig.~\ref{fig:model}(a), the effective model gives an adequate 
description of the lowest states of the Hubbard model. The obtained model 
parameters entering Eqs.~\eqref{Hmodel} and \eqref{kappaJ} are presented 
in Fig.~\ref{fig:model}(b) as functions of the interorbital hopping 
amplitude $\tilde{t}$. The realistic range of $E_T\approx 0.1-0.2\:\mathrm{eV}$ 
and $\kappa, J\approx 0.05-0.20\:\mathrm{eV}$ 
is obtained by taking $\tilde{t}\approx 0.2-0.4\:\mathrm{eV}$.
The small biquadratic exchange contained in $\mathcal{H}_\mathrm{eff}$ 
can be neglected at this point because the much larger effective 
biquadratic contribution is in fact generated by the $\kappa$-processes 
dynamically (see main text).

It is worth noticing that the strength $\kappa$ of the key process of the
model is finite due to interorbital hopping $\tilde{t}$. This process is 
thus ineffective in perovskite lattices, but it is perfectly allowed for the 
Fe-(As/Te)-Fe bonding geometry of Fe-pnictides/chalcogenides and leads to 
the spin-crossover mechanism (''soft'' magnetism) in these compounds 
(see main text). Concerning the role of intra-orbital $t$-hopping in the 
mapping, it did not enter the above formulas, since $t$ does not connect 
any pair of the selected low energy states. The intermediate states that 
can be reached by $t$ have an energy higher by $\Delta$ than those involved 
by $\tilde{t}$, so that the $t$-effect on $\kappa$ and $E_T$ values is 
relatively weak. It is only found to increase $J$ by about 
$2t^2/(\Delta+E_A-E_{TT})$. 

To conclude, we have shown that the model Hamiltonian proposed in the paper
naturally emerges from the two-orbital Hubbard model with strong Hund's 
coupling, when a regime of spin-state quasidegeneracy is realized. The model 
parameters that follow from this derivation are well within the ranges that 
we have explored in our study. 

\begin{thebibliography}{99}
\bibitem{note_orb} To address a tetra/ortho structural (''orbital order'') 
transition, we may include also $xz/yz$ orbital degeneracy of $S=1$ triplets; 
this is left for future work.
\bibitem{note_bqH} This can be understood using the identity 
$(\vc S_i\cdot\vc S_j)^2 = |\vc d_i\cdot\vc d_j|^2+1$. If $v\ll u\approx 1$, 
like in the FQ phase or close to it in the ns-AF phase, we recover the 
$\kappa$-term of Eq.~\eqref{eq:Hd}: $(\vc S_i\cdot\vc S_j)^2\approx 
\vc d^{\phantom{\dagger}}_i\!\cdot\!\vc d^{\phantom{\dagger}}_j 
+ \vc d^\dagger_i\!\cdot\!\vc d^\dagger_j$.
\end{thebibliography}
\end{document}